# EXCITATION SPECTRA OF CIRCULAR FEW-ELECTRON QUANTUM DOTS

L.P. Kouwenhoven, T.H. Oosterkamp, M.W.S. Danoesastro, M. Eto,

D.G. Austing, T. Honda, and S. Tarucha.

**Abstract.** We study ground states and excited states in semiconductor quantum dots containing 1 to 12 electrons. For the first time, it is possible to identify the quantum numbers of the states in the excitation spectra and make a direct comparison to exact calculations. A magnetic field induces transitions between excited states and ground state. These transitions are discussed in terms of crossings between single-particle states, singlet-triplet transitions, spin polarization, and Hund's rule. Our impurity-free quantum dots allow for atomic physics experiments in magnetic field regimes not accessible for atoms.

_________________________________________________________________

L.P. Kouwenhoven, T.H. Oosterkamp, M.W.S. Danoesastro, Department of Applied Physics and DIMES, Delft University of Technology, PO Box 5046, 2600 GA Delft, The Netherlands.

M. Eto, Department of Physics, Faculty of Science and Technology, Keio University, 3-14-1 Hiyoshi, Kohoku-ku, Yokohama 223, Japan.

D.G. Austing, T. Honda, and S. Tarucha, NTT Basic Research Laboratories, 3-1, Morinosoto Wakamiya, Atsugi-shi, Kanagawa 243-01, Japan.

_________________________________________________________________

**Introduction.** Small solid state devices known as quantum dots are often nicknamed *artificial atoms* since their electronic properties resemble the ionization energy and discrete excitation spectrum of atoms [1]. Quantum dots are usually fabricated between source and drain contacts such that the atom-like properties can be probed in current-voltage measurements. Additionally, with a nearby gate electrode, one can vary the number of electrons on the quantum dot, one-by-one, by changing the gate voltage and study any element in the "periodic table of artificial atoms". When an electron is added, the total charge on the dot changes by the elementary charge $e$. The associated energy change, known as the *addition energy*, is a combination of the single electron charging energy and the change in single-particle energy. Charging effects and discrete single-particle states have been studied in a variety of quantum dot systems, defined not only in semiconductors but also in metal grains and molecules [2].

Quantum dot devices usually contain some disorder, for instance, due to impurities or when the shape of the dot is irregular [2]. Clean quantum dots, in the form of regular disks, have only recently been fabricated in a semiconductor heterostructure [3,4]. The rotational symmetry of the two dimensional (2D) disks gives rise to a 2D shell structure in the addition energies, analogous to the 3D shell structure in atomic ionization energies [5]. From measurements of the *ground states* we have recently concluded [4] that the 2D shells in dots are filled according to Hund's rule: up to half shell filling, all electrons have parallel spins; more electrons can only be added with anti-parallel spins. Besides ground states, the other important energies in atomic-like structures are the *excitation energies*. In this Article we investigate the spectra of excitation energies for different numbers of electrons $N$ on the dot. For the first time, we are able to identify the quantum numbers of the states in the excitation spectra. In addition, we are able to show the relation between spectra of successive $N$ and how the spectra evolve with an applied magnetic field. The relatively large dimension of quantum dots (~ 100 nm) allows to use accessible magnetic fields to modify the excitation spectra, leading to the observation of transitions between states. Similar transitions in atoms require magnetic fields of the order $10^6$ T, several orders larger than that available in the laboratory.

**Device and experimental set-up.** Figure 1a shows a schematic drawing of the device which, from bottom to top, consists of an n-doped GaAs substrate, undoped layers of 7.5 nm $Al_{0.22}Ga_{0.78}As$, 12 nm $In_{0.05}Ga_{0.95}As$, 9.0 nm $Al_{0.22}Ga_{0.78}As$, and a ~500 nm n-doped GaAs top layer. A sub-micrometer pillar is fabricated using electron-beam lithography and etching techniques [3]. Source and drain electrical wires are connected to the top and substrate contacts. A third wire is attached to the metal that is "wrapped" around the pillar. This electrode is the side gate. The energy landscape along the vertical axis is shown in Figure 1b. Hatched regions denote occupied electron states in the conducting source and drain. The AlGaAs layers are insulating, but thin enough to allow for tunneling from the source to drain through the central InGaAs layer. The InGaAs layer has a disk-like shape. By making the gate voltage, $V_g$, more negative we can electrically "squeeze" the effective diameter of this disk from a few hundred nanometers down to zero. Concurrently, the number of electrons decreases one-by-one from about seventy to zero. At a particular gate voltage, i.e. for a particular electron number, we can spectroscopicaly probe the excitation spectrum by increasing the source-drain voltage, $V_{sd}$. Increasing $V_{sd}$ opens up a *transport window* between the Fermi energies of the source and drain. Those ground states and excited states lying within this energy window can contribute to current (see Figure 1b). When the gate voltage is increased the levels in Figure 1b shift down in energy. When an extra state moves through the Fermi energy of the drain the current increases. Changes in the current, $I$, are thus a direct measure of the energy spectrum of the dot. It is important to emphasize that in contrast to atoms, excitations do not occur inside the dot by, for instance, absorption of radiation. In the case of dots, excitations are created when an electron tunnels out from the ground state and the next electron tunnels in to an excited state. Our samples are measured while mounted in a dilution refrigerator with a temperature of 20 mK. Due to pick-up of noise the effective electron temperature is about 100 mK.

**Electron Transport Measurements.** Figure 2 shows the differential conductance $\partial I/\partial V_{sd}$ versus the source-drain voltage $V_{sd}$ on the horizontal axis and the gate voltage $V_g$ on the vertical axis. The electron number $N$ increases from 0 to 12. Along the vertical $V_{sd} \approx 0$ axis $N$ changes to $N+1$ when adjacent diamond-shaped regions of zero current touch. The size of the diamonds is a measure of the minimum energy to add or subtract an electron. As shown in Figure 2 the diamonds for $N$ = 2, 6 and 12 are unusually large - a consequence of having filled shells [4]. Also, the diamonds for $N$ = 4 and 9 are large relative to the neighboring diamonds. For these numbers, the second and third shell are half filled with parallel spins in accordance with Hund's rule [4]. These principal and secondary enhancements of the addition energies correspond directly to the large ionization energies found for the noble elements He, Ne, Ar, and enhanced ionization energies for N and P where the 2p and 3p shells are half filled with spin-polarized electrons [5].

At the two upper edges of the $N$ electron diamond an extra electron can tunnel through the dot via the $N+1$ electron ground state. Excited states of the $N+1$ electron system that enter the transport window are seen as "lines" running parallel to the diamond edges. Similarly, at the two lower edges of a diamond an electron can tunnel out of the dot, i.e. transitions between $N$ and $N$-1. However, some of the lines outside the diamonds in Figure 2 may be due to fluctuations in the number of states in the narrow leads [6,7]. Below we show that we can distinguish between lead and disk states by measuring $I(V_g,B)$ at different values for $V_{sd}$; i.e. the magnetic field, $B$, dependence of vertical "slices" through Figure 2.

Figure 3 shows the magnetic field dependence of the ground states. We have taken $V_{sd}$ = 0.1 mV which is sufficiently small that only ground states can lie within the transport window. The observed peaks in the current at $B = 0$ directly correspond to the touches of the $N = 0$ to $N = 5$ diamonds in Figure 2. The $B$-field dependence of the peak positions in gate voltage reflects the evolution of the ground state energies. Besides an overall smooth $B$-field dependence, we observe several kinks which we have labeled. For the regions between kinks, as we discuss below, we can identify the quantum numbers, including the spin configurations.

Increasing the source-drain voltage to $V_{sd} = 3$ mV yields the data summarized in Figure 4 [8]. This relatively large $V_{sd}$ increases the transport window such that the current is non-zero over wider gate voltage ranges. Instead of "sharp" current peaks as in Figure 3, we now observe "stripes". Adjacent stripes sometimes overlap implying that here $eV_{sd}$ exceeds the addition energy. The lower edge of the $N$th current stripe (which lies between the Coulomb blockade regions of $N$-1 and $N$ electrons) measures when the ground state of the $N$ electron dot enters the transport window as the gate voltage is made more positive. Inside a stripe we observe some random-looking and less-pronounced current changes which we attribute to fluctuations in the density of states in the leads [6]. However, also inside the stripes current changes can be seen as pronounced curves. As we argue below, these are the excited states in the dot.

For $N = 1$ a transport window of 3 meV is too small to observe clearly the excitations. Therefore, we show in Figure 5a the $N = 1$ stripe and a part of the $N = 2$ stripe for $V_{sd} = 5$ mV. For this voltage the $N = 1$ and 2 stripes just touch at $B = 0$. A pronounced current change from blue to dark red (i.e. from < 1 pA to > 10 pA) enters at the upper edge of the $N = 1$ stripe at $B = 0.2$ T. This change identifies the first excited state for the $N = 1$ dot (we discuss the index (0,1) below). Note that at higher $B$ values also higher excited states enter the stripe. The energy separation between ground state and first excited state can be read from the relative position inside the stripe. For instance, when the first excited state is 2/3$^{rd}$ of the width of the stripe away from the ground state the excitation energy is 2/3 times the source-drain voltage. So, the excitation energy is 5 meV at $B = 0$ and decreases for increasing $B$. Note that even over this wide magnetic field range of 16 T the first excited state never crosses with the ground state. Below 4 T, the excitation energy exceeds 3 meV and therefore the first excited state only starts to become visible for $B > 4$ T in the first stripe of Figure 4.

In the second, $N = 2$ stripe in Figure 4 we see the first excited state crossing with the ground state at $B = 4.15$ T; i.e. the first excited state for $B < 4.15$ T (seen as the current change from blue to red inside the second stripe) becomes the ground state for $B > 4.15$ T. So, the kink labeled by ▼ in Figure 3 is a crossing between the ground state and first excited state. For $N = 3$ and 4 we also observe a crossing at 1.7 T in the middle of the third and fourth stripes in Figure 4. This is a crossing between an up-going excited state with a down-going excited state. A similar up and-down going crossing is seen in the ground state for $N = 5$ at 1.2 T (see also the kink in Figure 3 labeled by ◆). We thus observe different types of crossings whose origins we now discuss.

**Discussion and comparison to theory.** To describe the electron states in a quantum dot we need to calculate the spectrum of energies $U(N,B)$ for a given number of electrons as a function of magnetic field. In our experiment we measure the electrochemical potential of the quantum dot which is defined as $\mu(N) \equiv U(N)-U(N-1)$; i.e. $\mu(N)$ is the *extra* energy necessary to add the $N$th electron. (Note that transitions in, for instance, $U(2)$ will be reflected in both $\mu(2)$ and $\mu(3)$.) For the case of a few electrons the energy spectrum can be calculated exactly, including the Coulomb interactions [9]. However, it is easier to explain the experimental results when we first consider the spectrum of non-interacting electrons in a 2D circular disk. The thickness of the thin disk freezes the electrons in the lowest state in the vertical direction.

We therefore only have to consider the confinement in the plane of the disk. For this we take a parabolic potential $V(r) = \frac{1}{2}m^*\omega_o^2 r^2$, where $m^* = 0.06 m_o$ is the effective mass of electrons in our InGaAs disk. The single-particle eigenenergies with radial quantum number $n = 0, 1, 2,...$. and angular momentum quantum number $l = 0, \pm 1, \pm 2,...$. are given by [10]:

$$E_{n\ell} = (2n + |\ell| + 1)\hbar\sqrt{(\tfrac{1}{4}\omega_c^2 + \omega_o^2)} - \tfrac{1}{2}\ell\hbar\omega_c, \tag{1}$$

where the cyclotron frequency $\omega_c = eB/m^*$. We neglect the much smaller Zeeman energy. The first few spin-degenerate states, $E_{n,l}$, are plotted in Figure 6a taking a confinement energy $\hbar\omega_o = 5$ meV. The two thick solid lines identify the transport window relative to the (0,0) curve for $V_{sd} = 5$ mV. The states within this stripe can be compared to the observed current changes seen in the $N = 1$ stripe in Figure 5a. The agreement is not unexpected since the non-interacting model of Eq. 1 is in fact exact for one electron on the dot. We note that Eq. 1 with $\hbar\omega_o = 5$ meV fits both the ground state and the first excited state very well up to about 7 T [11].

If we first neglect Coulomb interactions for an $N = 2$ dot then the two electron ground state energy is given by $U(N=2) = 2E_{0,0}$ and the measured electrochemical potential $\mu(2) = U(2) - U(1) = E_{0,0} = \mu(1)$. The simplest way to include interactions is to assume that the Coulomb energy $E_c$ is independent of $B$. In this constant interaction model, $\mu(2) - \mu(1) = E_c$; implying that the first and second current peaks are separated by a constant gate voltage and both peaks have the same $B$-field dependence. The constant interaction model has been successful in describing most Coulomb blockade experiments [1,2]. However, we see in Figs. 3 and 4 that the $N = 1$ and 2 ground states evolve differently with $B$. In particular, while $E_{0,0}$ is the $N = 1$ ground state over the entire $B$ range, a transition occurs at 4.15 T in the $N = 2$ ground state. To explain this transition we have to consider that the $l \geq 0$ orbits shrink in size when the magnetic field is turned on. Two electrons in a shrinking $l = 0$ orbit experience an increasing Coulomb interaction. (We indeed observe in Figure 3 that the second peak increases faster with $B$ compared to the first peak.) The increasing Coulomb interaction will, at some magnetic field, force one of the two electrons to occupy the larger $l = 1$ orbit. (We can identify the first excited state in the second stripe in Figure 4 as $E_{0,1}$ since it has almost the same $B$-field dependence as the first excited state in the $N = 1$ stripe in Figure 5.) This transition costs kinetic energy ($E_{0,1}$ - $E_{0,0}$), but it reduces the Coulomb interaction due to the larger spatial separation between the two electrons. In addition, the system gains exchange energy when the two electrons take on parallel spins. The transition in angular momentum is thus accompanied by a transition in the total spin from $S = 0$ (i.e. a singlet state) to $S = 1$ (i.e. a triplet state). An analogous singlet-triplet transition is expected to occur in helium atoms in the vicinity of white dwarfs and pulsars at $B = 4 \times 10^5$ T [12]. Due to the much larger size of our artificial atoms the transition occurs at accessible fields of a few Tesla. This was first predicted by Wagner et al. [13] and observations of kinks in the ground state energies were reported in Ref. [14].

Figure 6b shows an exact calculation of the electrochemical potential $\mu(N)$ for the $N = 2$ to 5 ground states and first two excited states (details of the calculation can be found in ref. 15). These exact spectra show extra transitions between many-body states that are not included in the single-particle states of Eq. 1. The singlet-triplet transition for $N = 2$ is one such example [16]. While in Figure 6a $E_{0,0}$ never crosses with $E_{0,1}$, Fig 6b shows a transition labeled by ▼ between the first (dashed) excited state and (solid) ground state at $\omega_c = 1.5\omega_o$. For $\hbar\omega_o = 5$ meV this singlet-triplet transition is expected at $B = 4.2$ T which is in remarkable agreement with the experimental value of $B = 4.15$ T in Figures. 3, 4, and 5a. Note that the calculated second (dotted) excited state in Figure 6b for $N = 2$ can also be seen in the second stripe of Figure 5a (i.e. the line between blue and red current regions which has a maximum near ~ 2 T).

We now discuss transitions between the first excited state and the ground state for $N = 3$, 4, and 5. The ground state for $N = 3$ has two transitions labeled by ●. On increasing $B$ the total spin, $S$, and total angular momentum, $M$, of the many-body states changes from $(S,M) = (½,1)$ to $(½,2)$ to $(^3/_2,3)$. These transitions to larger angular momentum states reduce the Coulomb interactions. In addition, the spin increases in order to gain exchange energy. A double transition in the ground state energy is indeed observed as two kinks in the $N = 3$ trace of Figure 3. In most regions in Figure 3 there is one main configuration for the occupation of single-particle states. For $N = 3$, in the region between the two labels ●, there are two important configurations, which both have the same total spin $S$ and total angular momentum $M$. In a similar way, the $N = 4$ and 5 ground states make transitions to higher angular momentum states and an increasing total spin when $B$ is increased. The occupation of many-body states in the region between the two labels ● is hard to determine since in this region different states lie very close in energy (see Figure 6b). For $B$ larger than the right ● there is again a clear ground state where electrons are fully spin-polarized and they occupy sequential angular momentum states. From the agreement between experiment and calculation we conclude that we have identified the quantum numbers and the transitions, labeled in Figure 3, in the ground states for $N = 2$ to 5.

A different type of crossing is between two excited states; i.e. crossings inside a stripe. We now argue that the crossing between two excited states in the $N = 3$ and 4 stripes labeled by ◆ in Fig 6b is a crossing between single-particle states. For non-interacting electrons we expect from Figure 6a that $E_{0,0}$ and $E_{0,1}$ are the two occupied single-particle states in the ground states for both $N = 3$ and 4. The first excited state is $E_{0,-1}$ for $B < 2$ T and $E_{0,2}$ for $B > 2$ T. Together with the ground state they form a triangle. The same triangular shape is observed in the $N = 3$ and 4 stripes in Figure 4 where it has a maximum near 1.7 T. (A closer look suggests that the crossing is not complete and that some level repulsion takes place.) Continuing these arguments we expect the transition in the first excited state for $N = 3$ and 4 to become a transition in the ground state for $N = 5$. Indeed this is seen in Figure 4 and at the kink labeled by ◆ in Figure 3 (note that we have indicated in Figure 3 this transition in $l$ from -1 to 2 in the diagrams adjacent to this kink). We emphasize that the discussion of the above crossings demonstrates for the first time a relation between the excitation spectrum of an $N$ electron system with the ground state of the $N+1$ electron system. We can thus conclude that for a few electron system the addition of an extra electron does *not* lead to complete mixing of the single-particle states [17].

The last crossing we discuss is indicated by ■ in Figure 3. We have identified this crossing earlier [4] as a manifestation of Hund's rule. As the adjacent spin diagrams show for the third and fourth electrons a transition from parallel spins (in accordance with Hund's rule) to anti-parallel spins occurs. When the states $E_{0,1}$ and $E_{0,-1}$ are sufficiently close, there is an energy gain due to the exchange interaction between electrons with parallel spins [5]. As $B$ is increased, $E_{0,1}$ and $E_{0,-1}$ diverge from each other (see Figure 6a) and at some value a transition is made to anti-parallel spins where the third and fourth electrons both occupy $E_{0,1}$. Figure 5b shows a surface plot of the $N = 4$ stripe measured at $V_{sd} = 1.6$ mV. This surface plot clearly shows the magnetic field dependence of the single-particle states $E_{0,1}$ and $E_{0,-1}$ including a Hund's rule crossing between the ground state and first excited state at 0.4 T. Interestingly, a *second* excited state is seen with a $B$ dependence parallel to the first excited state. Parallel first and second excited states are also seen in the calculation of Figure 6b (see just above and below the label ■ in the $N = 4$ stripe). Parallel $B$ dependence means that the many-body states have the same $M$. The difference between the two parallel lines is that in the lower energy line the third and fourth electrons have parallel spins (in accordance with Hund's rule) and in the higher energy line they have anti-parallel spins. The energy difference is a direct measure of the gain in exchange energy. From the experimental $N = 4$ stripe in Figure 4 we can read directly that the gain in exchange energy is ~1 meV.

In conclusion we have studied the excitation spectra of few electron quantum dots. By measuring the magnetic field dependence of the current peaks at different source-drain voltages it is possible to identify the quantum numbers of the electronic states. For the first time, a comparison is made to calculated excitation spectra. The results have been reproduced in different samples which implies that it is now possible to fabricate identical solid state devices on the level of single electron states.

**Figure captions.**

Figure 1. (a) Schematic drawing of the semiconductor layers and the metal side gate. The diameter of the pillar is 0.5 µm. (b) Schematic energy (horizontal axis) diagram along the vertical axis of the pillar. Hatched regions are occupied electron states in the source and drain contacts. For the case shown, two electrons are permanently trapped in the quantum dot. The third electron can choose to tunnel through the $N = 3$ ground state (solid line) or through one of the two excited states which lie in the transport window. This situation corresponds to the $N = 3$ current stripe.

Figure 2. Differential conductance $\partial I/\partial V_{sd}$ plotted in color scale in the $V_g$ - $V_{sd}$ plane at $B = 0$. In the white diamond shaped regions $\partial I/\partial V_{sd} \approx 0$ due to Coulomb blockade. The number of electrons $N$ is fixed in the diamond regions. The lines outside the diamonds running parallel to the sides, identify excited states in the dot or leads.

Figure 3. $I(V_g,B)$ for $N = 0$ to 5 measured with small $V_{sd} = 0.1$ meV such that only ground states contribute to the current. Ground state transitions are indicated by different labels. The arrows in the squares indicate the spin configuration. The lowest square corresponds to a single-particle state with angular momentum $l = 0$. For squares to the right $l$ increases to 1, 2, 3, etc. For $N = 4$ and 5, near $B = 0$, also the $l = -1$ square is shown on the left of the $l = 0$ square. For $N = 3$ there are two important configurations for the occupation of single-particle states in the region between the two kinks.

Figure 4. $I(V_g,B)$ for $N = 0$ to 4 and a part of $N = 5$ measured with $V_{sd} = 3$ meV. $I < 0.1$ pA in the dark blue regions and $I > 10$ pA in the dark red regions. Now both ground states and the first few excited states can contribute to the current. Current stripes between the Coulomb blockade regions (i.e. dark blue) for $N$-1 and $N$ electrons are called the $N$ electron stripe throughout the paper.

Figure 5. (a) $I(V_g,B)$ for $N = 1$ and 2 measured with $V_{sd} = 5$ meV up to 16 T. The states in the $N = 1$ stripe are indexed by the quantum numbers $(n,l)$. (b) Surface plot of the $N = 4$ stripe measured with $V_{sd} = 1.6$ meV up to 2 T.

Figure 6. (a) Calculated energy spectrum from Eq. 1 for $N = 1$ and $\hbar\omega_o = 5$ meV. The lowest thick line is the ground state energy. The upper thick line is the ground state energy shifted upwards by 5 meV. Dashed states between the two thick lines can be seen in the experimental stripe for $N = 1$ in Figure 5. (b) Exact calculation of energy spectra for $N = 2$ to 5. Current stripes of width $0.66 \cdot \hbar\omega_o$ are bounded by solid lines. For $\hbar\omega_o = 5$ meV, $\omega_c = \omega_o$ corresponds to 2.8 T. The labels ■, ●, ▼, and ◆ indicate the same transitions as in Figure 3.

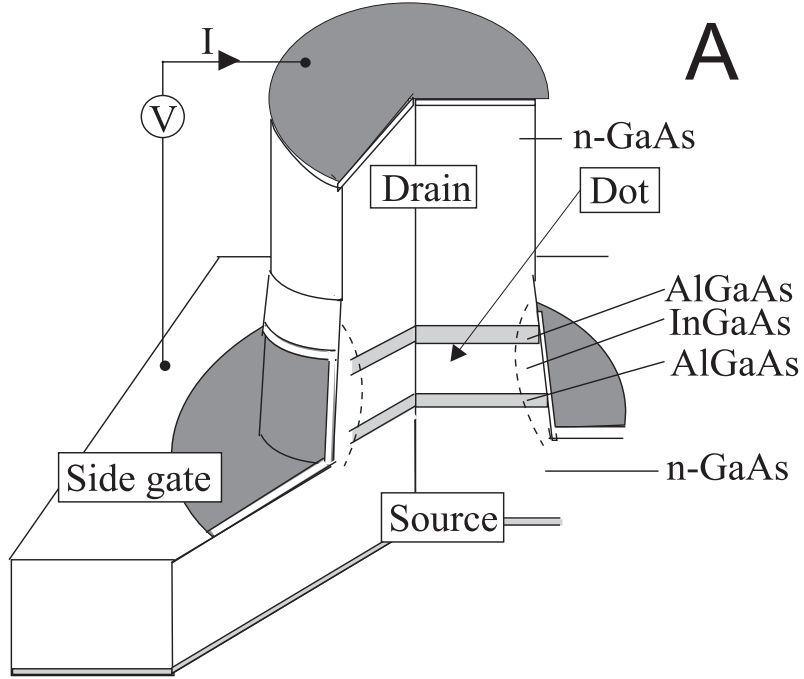
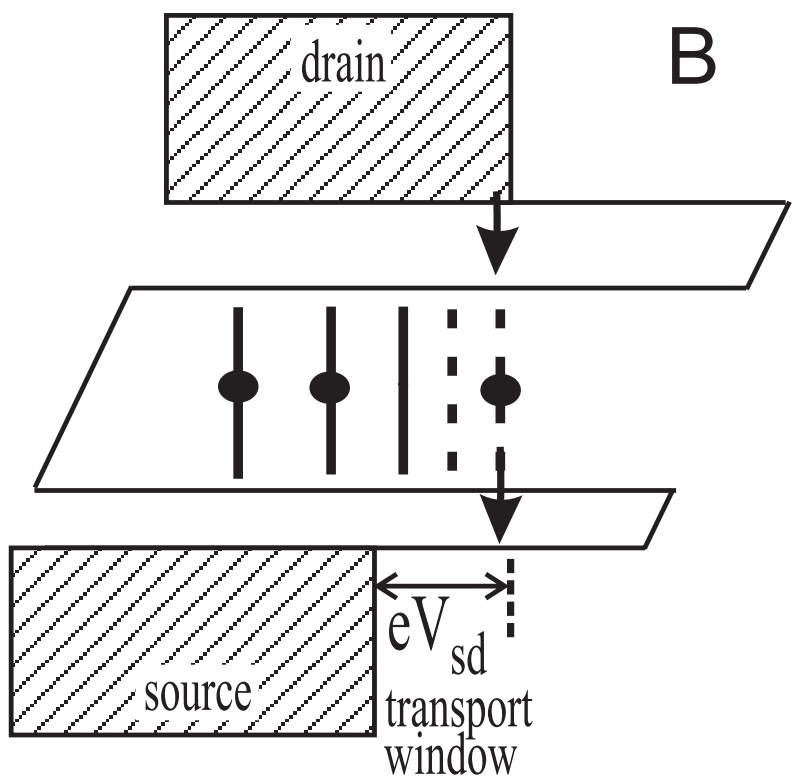

Figure 1. Kouwenhoven et al.

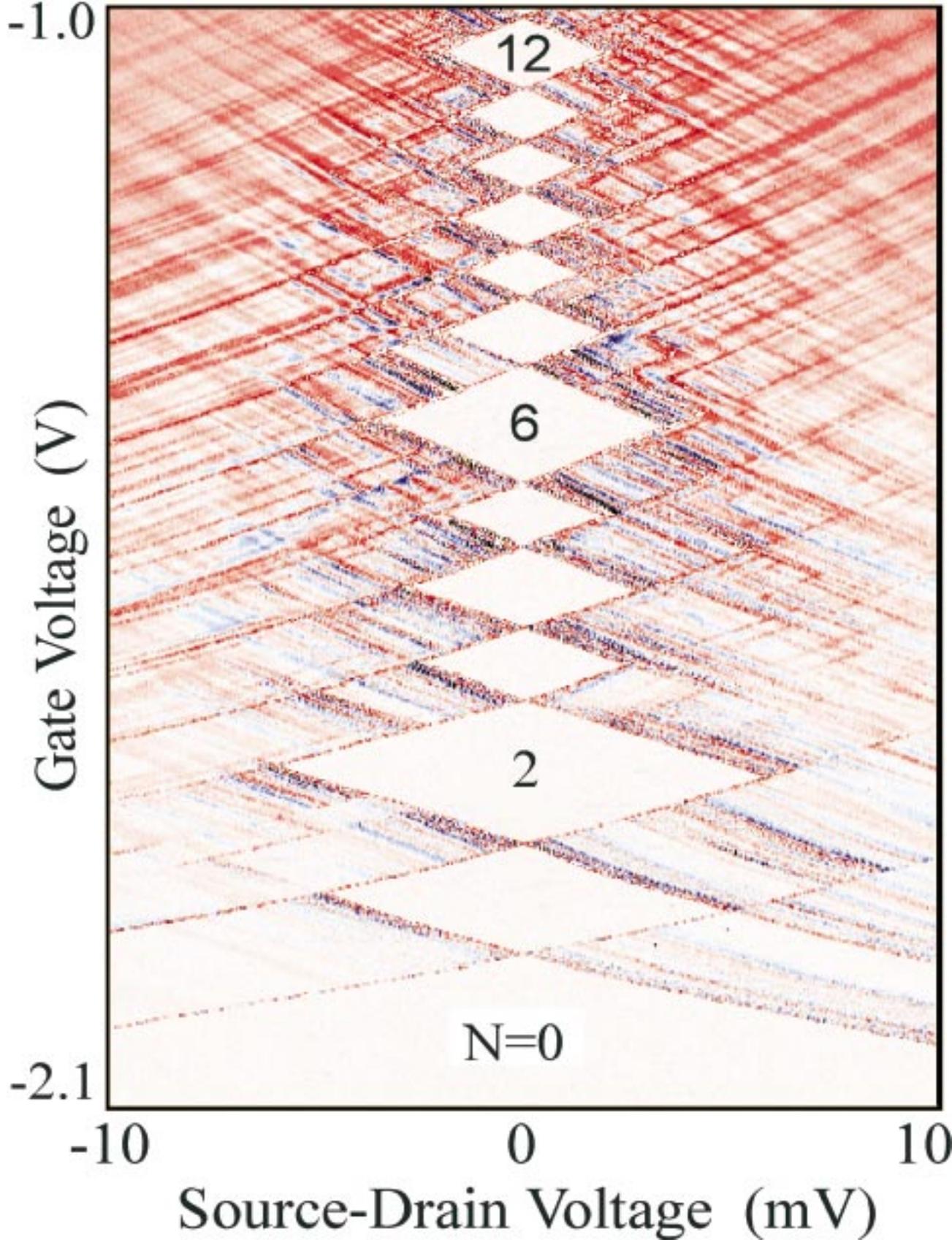

Figure 2

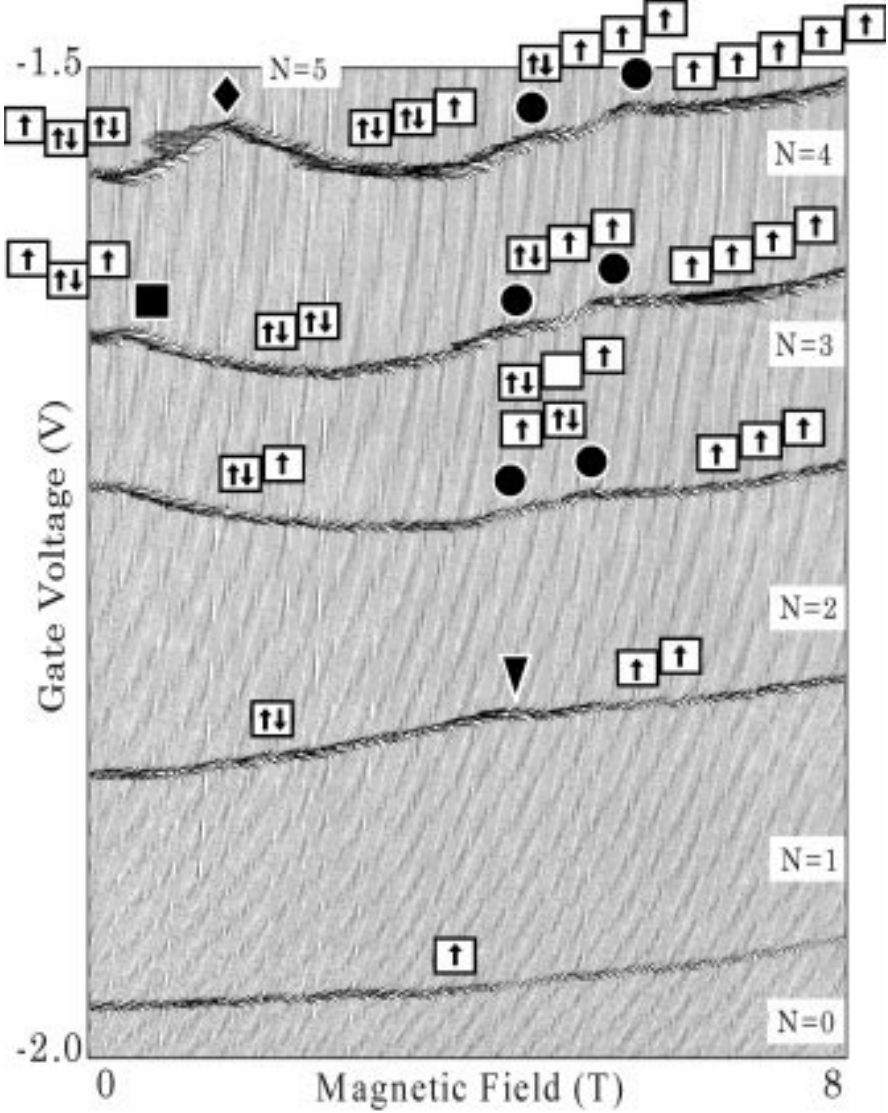

Figure 3

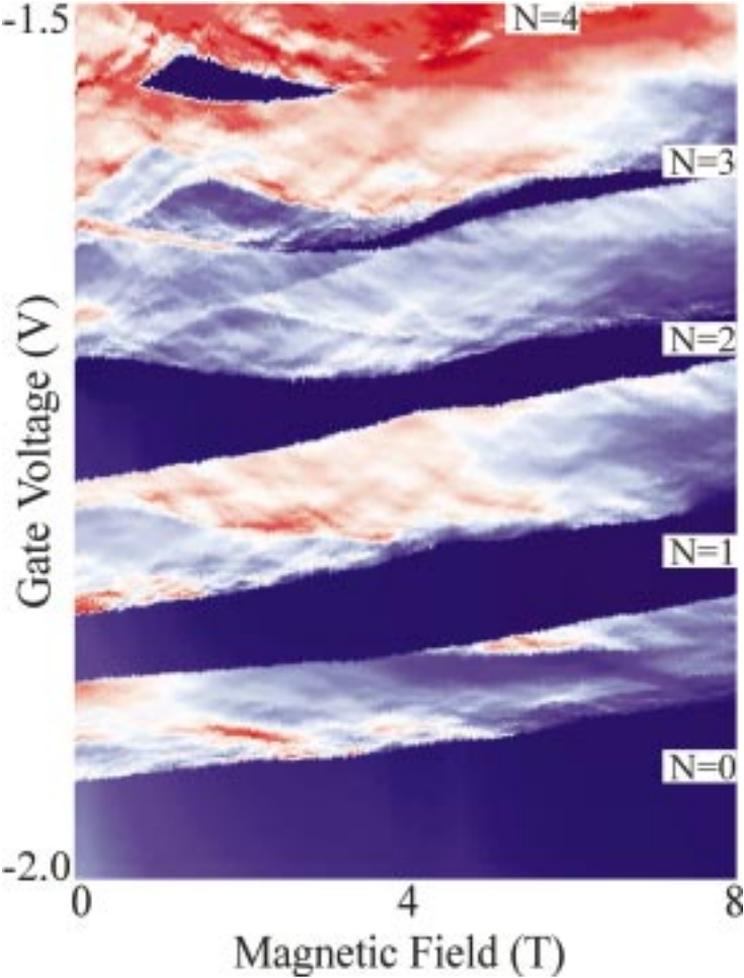

Figure 4

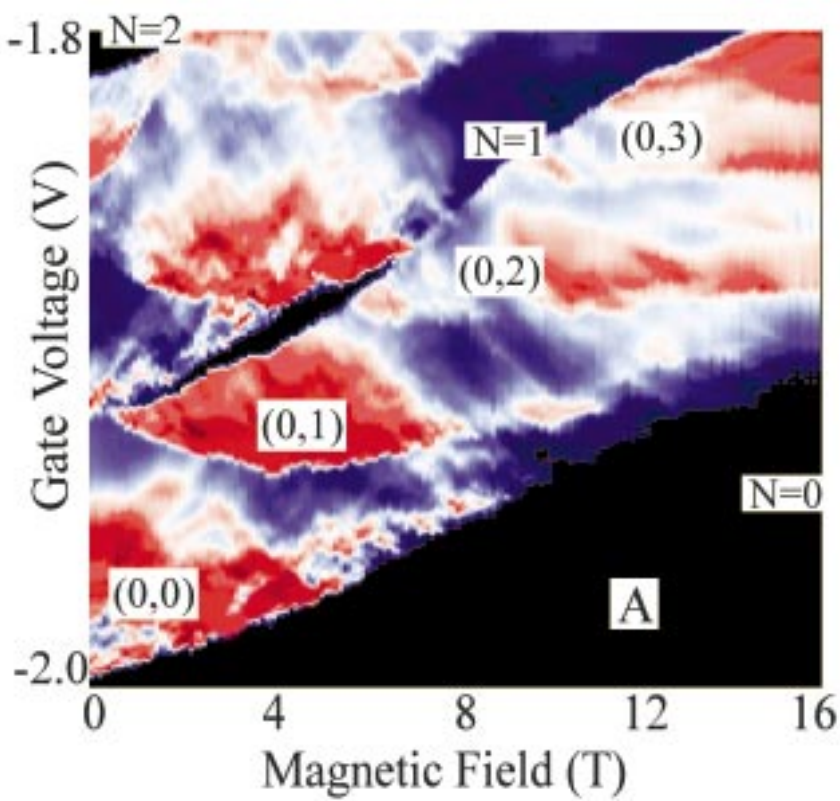
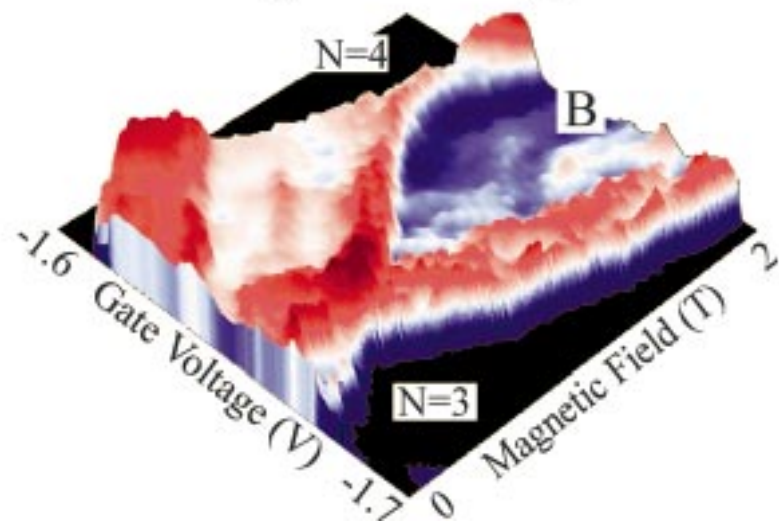

Figure 5

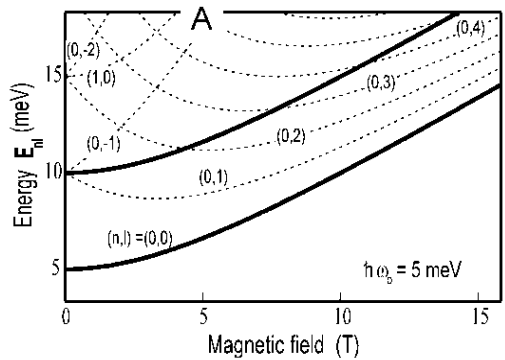
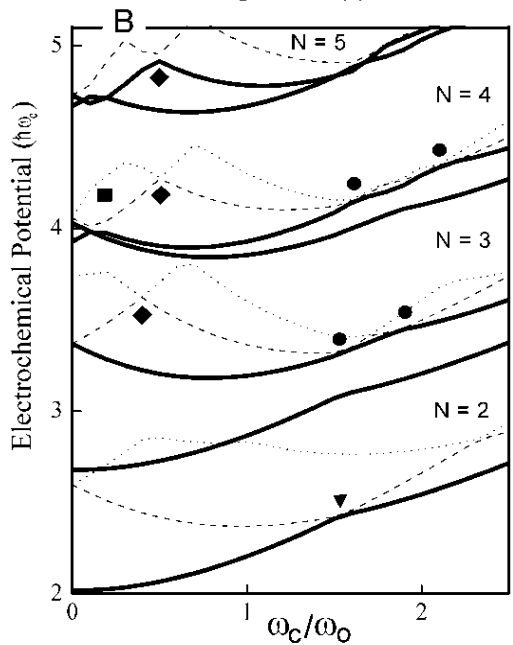

Figure 6  Kouwenhoven et al.